# A Study of End-Fed Arrays with Application to Single Driving-Point Self-Standing Linear Arrays

*Konstantinos KONDYLIS*[1], *Nikolitsa YANNOPOULOU*[1], *Petros ZIMOURTOPOULOS*[2]

[1]Antennas Research Group, Palaia Morsini, Xanthi, Thrace, Hellas, EU
[2]Antennas Research Group, Department of Electrical Engineering and Computer Engineering, Democritus University of Thrace, V. Sofias 12, 671 00, Xanthi, Greece

k8k@antennas.gr, pez@antennas.gr, yin@antennas.gr, www.antennas.gr

**Abstract.** *The array factor of a both geometrically and electrically uniform array is the simple formula for the complex geometric progression sum. This fact, although results in the simplest of all possible analytical designs, obviously does not in the least simplify the complicated practical problem of feeding the array elements using multiple driving points. In order to begin the examination of uniform linear arrays with a single driving point, this short paper presents a compact study of the end-fed space arrays with application to geometrically uniform, self-standing linear arrays of parallel dipoles. A number of test array models were simulated, constructed and their radiation pattern was then measured. The experimental and computational results were found to be in good agreement. The developed software applications are available through the Internet as FLOSS Free Libre Open Source Software.*

## 1. Introduction

A space array of $1 \le k \le N$ parallel, arbitrarily shaped, dipoles with identical current density distributions, has a complex vector radiation pattern $E = AG$ (by applying the "Principle of Radiation Patterns Multiplication", i.e. by the linear property of the volume integral), where $G$ is the radiation pattern of the first dipole (which "generates the array"; the "Generator Pattern"), and $A$ is the Array Factor,

$$A = \sum_{k=1}^{N} \left(\frac{I_k}{I_1}\right) e^{i\beta(R_{k_r} - R_{1_r})} \qquad (1)$$

i.e. the complex numerical radiation pattern of $N$ invented "isotropic point sources", each of current $I_k$ and pointed by the dipole center vector $R_k$, with projection $R_{kr}$ to the unit direction vector $r$. Linear Arrays have dipole centers on a straight line. Fully Uniform Linear Arrays are both geometrically uniform, i.e. the dipoles are equidistant, and electrically uniform, i.e. the consecutive dipole currents are of equal amplitude and constant phase difference [1].

## 2. End-Fed Arrays

Perhaps, the simplest practical array is the one constructed from a linear two-wire transmission line that supports the arms of parallel, linear, symmetrical dipoles, vertical to its plane. In order to operate the line as a balanced one, it is end-fed through a coaxial line balun from a coaxial connector, which at the same time supports the weight of the whole, self-standing, array. A possible analysis of such an array is shown in Fig. 1.

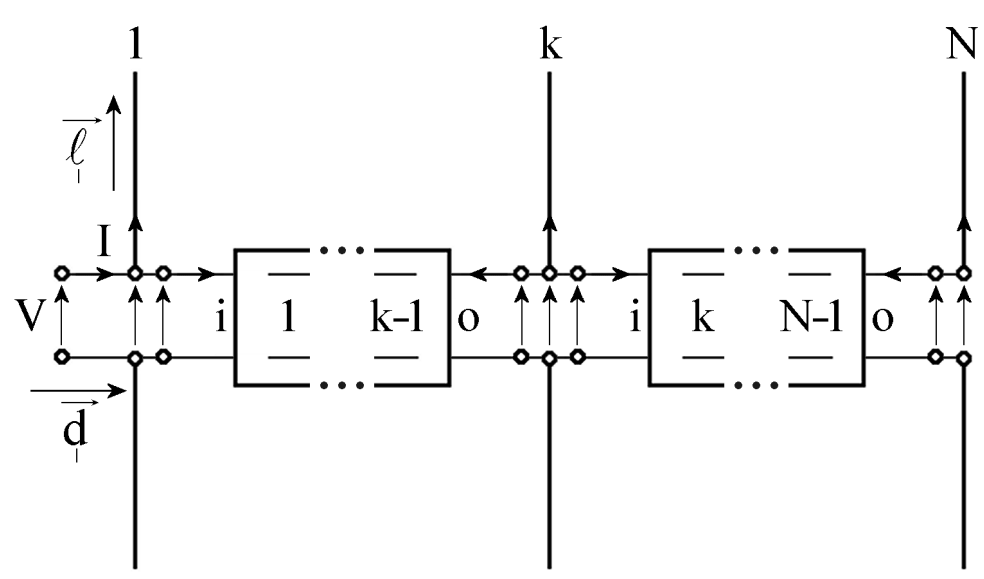

**Fig. 1.** End-fed linear array of linear dipoles.

In this one port linear network, the number of circuit voltage and current variables is equal to the sum of 2 variables for the input port plus $2N$ variables for the dipole ports plus $4(N-1)$ variables for the line segment ports, that is a total number of $6N-2$ variables. The number of the linear relations between these variables is equal to the sum of 3 relations, for the generator dipole

$$1 \le N : V = V_1 = {}_iV_1,\ I = I_1 + {}_iI_1, \qquad (2)$$

plus 2 relations, for the last dipole

$$2 \le N : {}_oV_{N-1} = V_N,\ {}_oI_{N-1} + I_N = 0, \qquad (3)$$

plus $3(N-2)$ relations, for the intermediate dipoles

$$3 \le N, 1 \le k \le N-2 : \quad {}_oV_k = V_{k+1} = {}_iV_{k+1},\ {}_oI_k + I_{k+1} + {}_iI_{k+1} = 0 \qquad (4)$$

plus $N$ relations, for the coupling between these dipoles

$$1 \le k \le N : V_k = \sum_{\mu=1}^{N} z_{k\mu} I_\mu \qquad (5)$$

plus 2($N$-1) relations, for each one of the lossless transmission line segments of electrical length $\beta l_k$ and characteristic impedance $Z_{0k}$

$$0<\beta\ell_k\neq\nu\pi,\ \nu=1,2,3,\ldots:$$

$$\begin{bmatrix} {}_{ii}z_k & {}_{io}z_k \\ {}_{oi}z_k & {}_{oo}z_k \end{bmatrix} = -i\,\frac{Z_{0k}}{\sin\beta\ell_k}\begin{bmatrix} \cos\beta\ell_k & 1 \\ 1 & \cos\beta\ell_k \end{bmatrix} \qquad (6)$$

$$\beta\ell_k=\nu\pi,\ \nu=2\mu+1,\ \mu=0,1,2,\ldots:$$
$${}_oV_k=-{}_iV_k\ ,\ {}_oI_k-{}_iI_k=0 \qquad (7)$$

$$\beta\ell_k=\nu\pi,\ \nu=2\mu+2,\ \mu=0,1,2,\ldots:$$
$${}_oV_k={}_iV_k,\ {}_oI_k+{}_iI_k=0 \qquad (8)$$

that is a total of 6$N$-3 linear relations to solve for 6$N$-2 variables, in terms of the source voltage $V$, which is considered as a parameter. Anyhow, the resulting current ratios are clearly independent of $V$.

# 3. Single Driving-Point Self-Standing Linear Arrays

The authors' group has limited available technical resources for antenna construction. This fact restricts the experimentation to thin-wire self-standing array models of a low total weight that is of a small total transmission line length and of a small number of dipoles. Thus, the practical application of the analysis was carried-out for $N$ = 2, 3 and 4 dipoles only. In order to demonstrate the procedure in use, the smallest case of $N$ = 2 dipoles is presented in some detail. In Fig. 2, the resulting 6$N$-3 = 9 linear relations between 6$N$-2 = 9 variables + 1 parameter, are shown in a compact form, for the case of $\beta l_1 \neq \nu\pi$, $\nu$ = 1, 2, ... in (6), with each cell value to be the coefficient of the variable in the first row of its column in an implied summation, while, if $\beta l_1 = \nu\pi$, $\nu$ = 1, 2, ... then the two last rows with the gray background have to substituted by the rows of Fig. 3 or 4, according to the odd or even value of $\nu$ given in (7)-(8).

| $V_1$ | $V_2$ | $I_1$ | $I_2$ | ${}_iV_1$ | ${}_oV_1$ | ${}_iI_1$ | ${}_oI_1$ | $I$ | = | $V$ |
|---|---|---|---|---|---|---|---|---|---|---|
| 1 | | | | | | | | | = | 1 |
| 1 | | | | −1 | | | | | = | 0 |
| | | 1 | | | | 1 | | −1 | = | 0 |
| | 1 | | | | −1 | | | | = | 0 |
| | | | 1 | | | | 1 | | = | 0 |
| −1 | | $z_{11}$ | $z_{12}$ | | | | | | = | 0 |
| | −1 | $z_{21}$ | $z_{22}$ | | | | | | = | 0 |
| | | | | −1 | | ${}_{ii}z_1$ | ${}_{io}z_1$ | | = | 0 |
| | | | | | −1 | ${}_{oi}z_1$ | ${}_{oo}z_1$ | | = | 0 |

**Fig. 2.** The linear system of 9 relations for arrays with $N$ = 2 and $\beta l_1 \neq \nu\pi$, $\nu$ = 1, 2, ... .

Obviously, the complexity of the expressions increase with the number of dipoles, from the simple, of 2 dipoles

$$I_{21}=\frac{I_2}{I_1}=\frac{z_{12}\,{}_{oo}z_1-{}_{io}z_1\,z_{22}}{{}_{io}z_1^2+{}_{io}z_1\,z_{12}-{}_{oo}z_1\left(z_{22}+{}_{oo}z_1\right)} \qquad (9)$$

in which the equality of self and mutual impedances have been taken into account, resulting from the system of Fig. 2, or of the simplest $I_2/I_1$ = +1 or $I_2/I_1$ = -1 of Fig. 3 or Fig. 4 respectively, to the most complex one for 4 dipoles, shown in Fig. 5, which covered about one and a half A4 page.

| | | | | | | | | | | |
|---|---|---|---|---|---|---|---|---|---|---|
| | | | | 1 | 1 | | | | = | 0 |
| | | | | | | −1 | 1 | | = | 0 |

**Fig. 3.** Last rows replacement for $\nu = 2\mu + 1$, $\mu$ = 0, 1, 2, ... .

| | | | | | | | | | | |
|---|---|---|---|---|---|---|---|---|---|---|
| | | | | −1 | 1 | | | | = | 0 |
| | | | | | | 1 | 1 | | = | 0 |

**Fig. 4.** Last rows replacement for $\nu = 2\mu + 2$, $\mu$ = 0, 1, 2, ... .

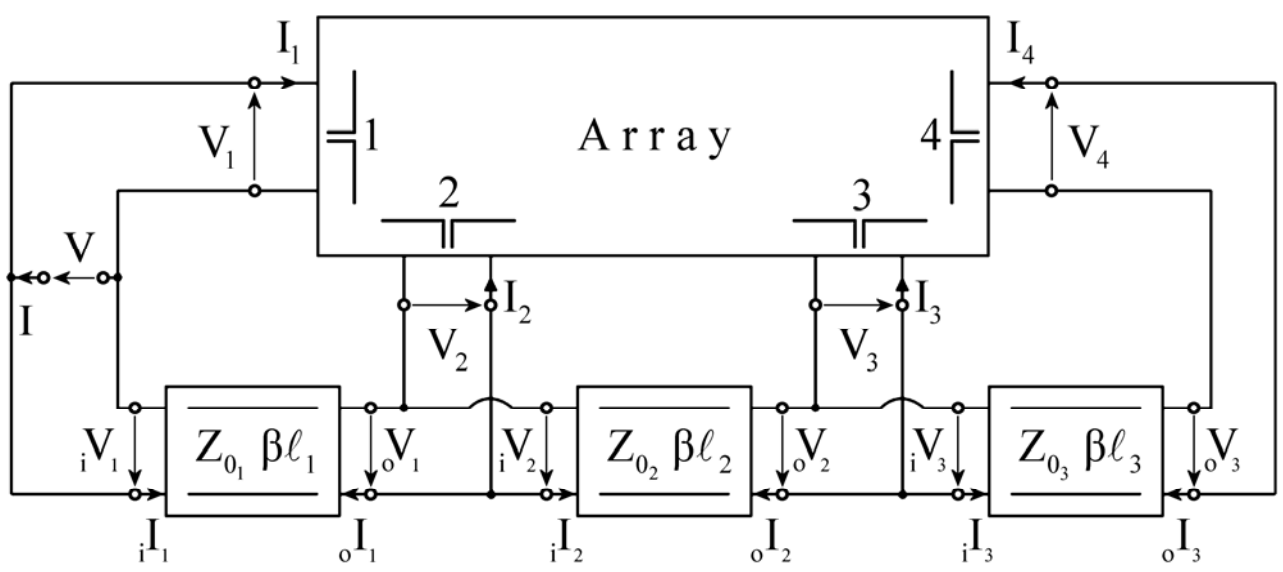

**Fig. 5.** Equivalent circuit of the 4 dipoles linear array.

Three Visual Fortran applications were developed for the computation of current ratios. The GUI application form for $N$ = 4 dipoles, is shown in Fig. 6 [2], [3].

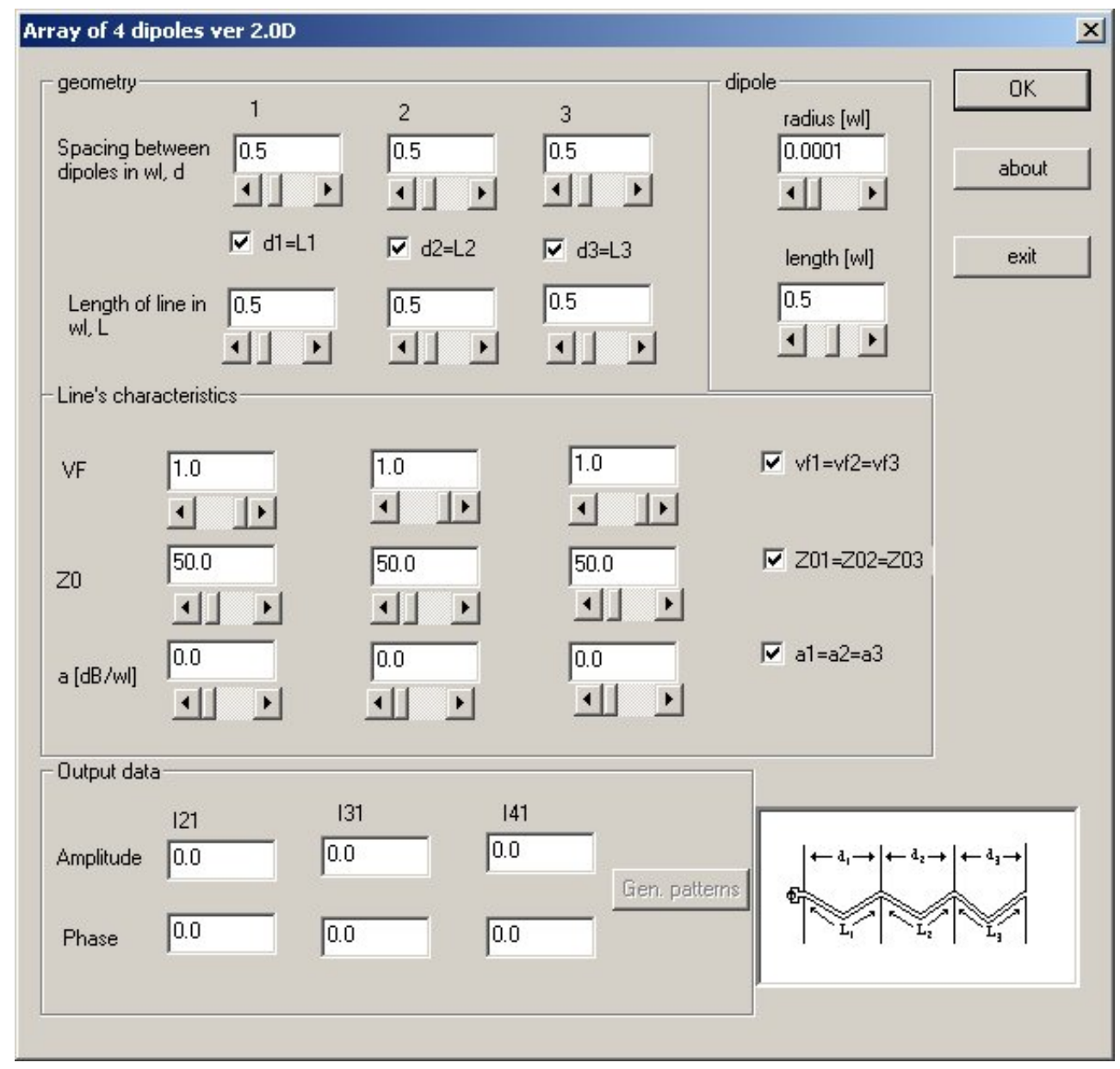

**Fig. 6.** GUI for the analysis of a linear array of $N$ = 4 dipoles.

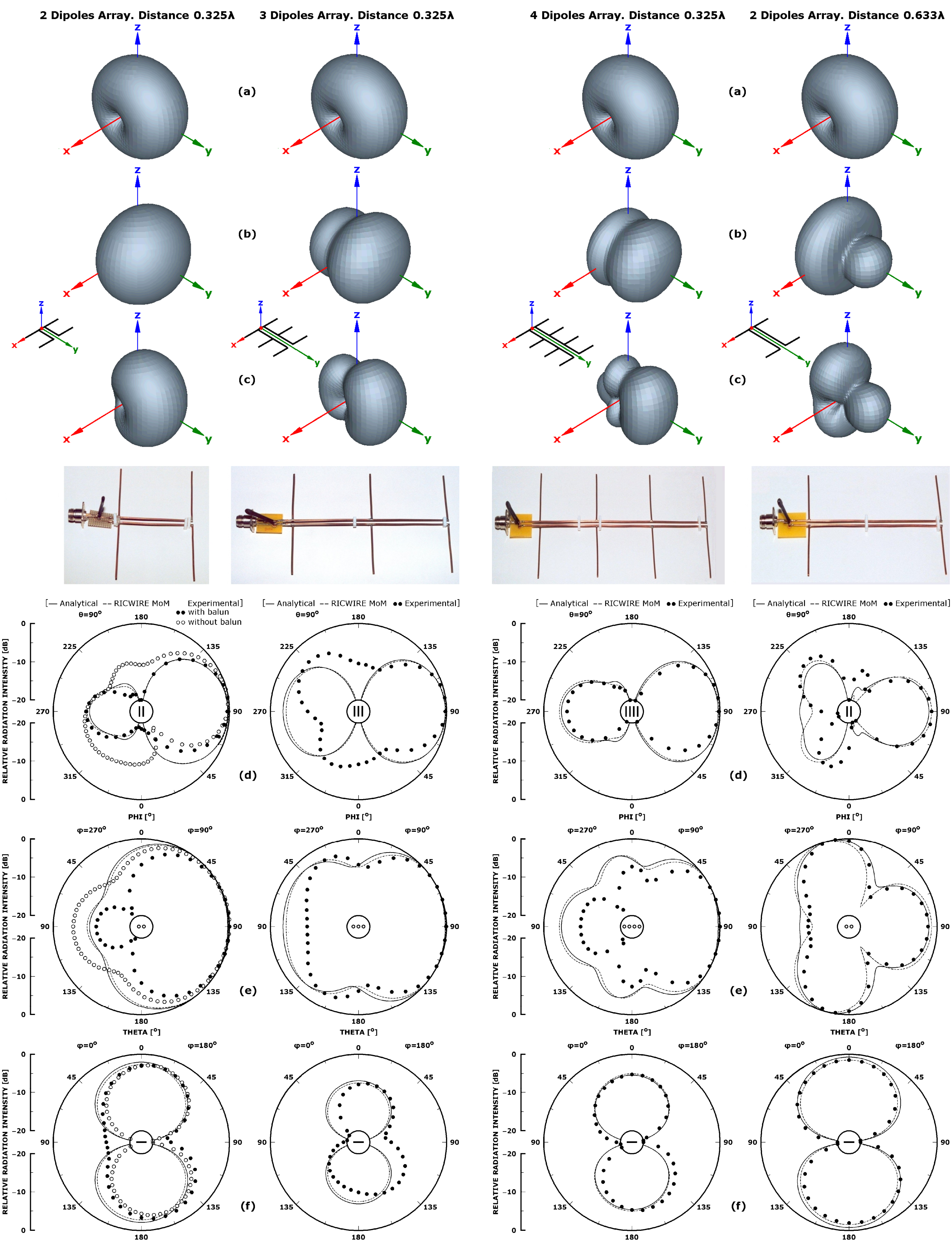

**Fig. 7.** Analysis, simulation and measurements for experimental arrays: $N = 2$, $N = 3$ and $d = 0.325\lambda$.

**Fig. 8.** Analysis, simulation and measurements for experimental arrays: $N = 4$, $d = 0.325\lambda$ and $N$=2, $d = 0.633\lambda$.

In this form, the input data are the $N$-1 distances between dipoles, the dipole radius and length, the length, the characteristic impedance and the velocity factor of each transmission line segment. In addition to current ratios, each application produces the text files needed by the [RadPat4W] application of a FLOSS mini-Suite of tools, which plots in Virtual Reality the Generator, Array Factor, and Array 3D radiation pattern, as well as, their 2D main-plane cuts [4]. The deduced formulas for the determination of the 1 + 2 + 3 = 6 current ratios ($I_k/I_1$), were mechanically verified using Mathematica.

# 4. Array Design, Construction and Measurements

A number of dipole array designs were carried out using the developed applications and the [RichWire] simulation program, which is a fully analyzed, corrected and redeveloped edition of the original Moment Method thin-wire computer program [4], [5]. Eleven arrays were finally constructed and tested. The results for the current ratios of four selected arrays are shown in Tab. 1. The arrays were designed for operation at the frequency of 1.111 GHz. A two-wire transmission line, of $Z_0$ = 200 Ω and velocity factor $vf$ = 1 was constructed to feed the dipoles [6]. This balanced line was then connected to an unbalanced 50 Ω type-N/F base connector through a 4:1 balun made from a segment of RG-174U coaxial cable ($Z_0$ = 50 Ω, $vf$ = 0.66) with total length λ/2.

| Dipoles | 2 | 3 | | 4 | | | 2 |
|---|---|---|---|---|---|---|---|
| L/λ | 0.5185 | 0.5 | | 0.5 | | | 0.44 |
| d/λ | 0.325 | 0.325 | | 0.325 | | | 0.633 |
| Ratios | $I_2/I_1$ | $I_2/I_1$ | $I_3/I_1$ | $I_2/I_1$ | $I_3/I_1$ | $I_4/I_1$ | $I_2/I_1$ |
| $\lvert I_k/I_1 \rvert$ | 0.534 | 0.205 | 0.286 | 0.280 | 0.115 | 0.226 | 0.527 |
| $\angle I_k/I_1$[°] | −80.5 | −114.9 | 173 | −115.9 | 150.4 | 68.1 | 49.5 |

**Tab. 1.** The experimental array characteristics and current ratios.

The arrays were constructed by bare copper wire of 1 mm (0.0037λ) radius and they are self-standing using an orthogonal piece of a two-sided printed board (3 cm x 4.48 cm), on which the two-conductor line was soldered. A few Teflon spacers of low relative dielectric constant (≅ 2) were fabricated to fix the distance between the two line wires at 5.5 mm. The measurement system consists of a 50 Ω Vector Network Analyzer external to an anechoic chamber [7]. Each array under test was azimuthally rotated around its three main axes, by a 360° built positioner, under the developed software control of a built hardware controller. The stationary antenna was a UHF standard gain antenna [8]. Fig. 7 and Fig. 8 show the results for the 4 test arrays, in 3 groups of rows, as follows: 1st group: The screen captures of the produced Virtual Reality radiation patterns in dB for (a) Generator, (b) Array Factor, and (c) Dipole Array - 2nd group: The constructed model, and 3rd group: The 2D radiation pattern cuts by (d) xOy, (e) yOz, and (f) zOx main-planes. In Fig. 7, the measurements for the array of $N$ = 2, dipoles with equidistance $d$ = 0.325λ, were carried out with and without balun. In Fig. 8, the array of $N$ = 2 dipoles with equidistance $d$ = 0.633λ, has been designed to exhibit the maximum radiation pattern direction off the 3 main-axes.

# 5. Conclusions

Although, the unavoidable mechanical supporting elements which exist in the anechoic chamber in the neighborhood of the antenna under test may affect the radiation pattern measurements, the observed differences in the 4 of the 12 patterns between the analysis and simulation results, on the one hand, and the measurements, on the other, have to be charged respectively: (1) On a cone-cut of the array radiation pattern, instead of the expected yOz main-plane cut, in Fig. 7(e), $N$=2 and Fig. 8(e), $N$=4, (2) On an inclination of the rotation axis relative to the expected linear polarization measurement plane, in Fig. 8(d), $N$=2, and (3) On a loosed connection during the array rotation, in Fig. 7(d), $N$=3. These conclusions are amplified by the careful study of the corresponding Virtual Reality space radiation patterns. Therefore, under the given measurement circumstances, the experimental and computational results were found to be in good agreement and no attempt was made to modify any design or repeat any measurement.

The results for the single driving-point self-standing fully Uniform Linear Arrays, i.e. those including the electrical uniformity, along with their application to the constrained pattern design will be presented in a future paper.